\documentclass[10pt ]{revtex4}
\usepackage{hyperref}
\usepackage{amsmath,amsfonts}
\usepackage[dvips]{graphics}
\usepackage[dvips]{graphicx}
\usepackage{subfigure}
\usepackage{epsfig}

\RequirePackage{color}

\begin{document}

\title{ Thermodynamics  and phase transitions of   galactic clustering in higher-order Modified Gravity  }

 \author{Sudhaker Upadhyay${}^{a}$}
  \email{sudhakerupadhyay@gmail.com}
  \author{Behnam Pourhassan${}^{b}$}
  \email{b.pourhassan@du.ac.ir}
 \author{Salvatore Capozziello${}^{c}$}
   \email{capozziello@na.infn.it}

 \affiliation{${}^{a}$Department of Physics, K.L.S. College Nawada, Nawada-805110, Bihar,  India}
  \affiliation{${}^{a}$Visiting Associate, Inter-University Centre for Astronomy and Astrophysics  (IUCAA) Pune, Maharashtra-411007}
 \affiliation{${}^{b}$School of Physics, Damghan University, Damghan, 3671641167, Iran}
 \affiliation{${}^{c}$Dipartimento di
 Fisica ``E. Pancini",  Universit\'a di Napoli  Federico II, I-80126 - Napoli,
  Italy. \\
  INFN Sez. di Napoli, Compl. Univ. di Monte S. Angelo,
 Edificio G, I-80126 - Napoli, Italy.\\   Gran Sasso Science
 Institute, Viale F. Crispi, 7, I-67100, L'Aquila, Italy.\\
 Tomsk State Pedagogical University, ul. Kievskaya, 60, 634061 Tomsk, Russia.}

\begin{abstract}
 We study the thermodynamics of galactic clustering  under the higher-order corrected Newtonian dynamics.  The clustering of galaxies  is considered as a
 gravitational phase transition. In order to study the effects of higher-order correction
 to the thermodynamics of gravitational system, we compute more exact equations of state.
 Moreover, we  investigate the corrected
 probability distribution function for such gravitating system. A relation between
 order parameter
 and the critical temperature  is also established.
\end{abstract}
\maketitle

\section{Overview and Motivation}
Because of gravitational interaction of galaxies, the characterization of galactic clustering is a subject of extensive interest. It is well known that the gravitational interaction between clusters of galaxies plays an important role in the evolution of universe.
A mathematical description which explains the current behavior of universe and its evolution over time is known as the cosmological model. The considered cosmological model in this paper is based on the fundamental assumptions which may explain  numerous   observations.
Though it is clear that there is some missing mass in the universe which is known as the dark matter. This is because of the gravitational effects on the clusters of galaxies.
There are some reports available on  expected discoveries
of dark matter \cite{ber}, while  direct observation for the dark matter does not exist.
Dark matter doesn't interact electromagnetically which make it impossible to detect.
A less explored
route is that our current theory explaining universe  might
be incomplete. Milgrom has
proposed an empirically motivated modification   of Newtonian dynamics  at low accelerations, known
as MOND \cite{mil}, which  has been found extremely successful
in explaining some observational properties
of galaxies \cite{san22}. Yang et. al. shows how modification of the Newtonian potential would modify the gravitational effects on the statistical mechanics \cite{yang} and it was the first analysis of clustering with modifications of Newtonian gravity. Also, if the MOND approach
is not covariant and hence cannot be studied
in a general setting, it is possible to show that this can be framed in the weak field limit of $f(R )$  gravity \cite{bernal}.
A relativistic MOND
 theory is explored which resolves the  problems of  gravitational lensing,   violated hallowed principles by
exhibiting superluminal scalar waves or an a priori vector field \cite{beke}.
MOND also predicts the detailed shape of a rotation curve from the observed matter   distribution successfully \cite{kent,milo,beg}.  In fact,   many theoretical model and indirect study are devoted to understanding the nature of the dark matter \cite{ell,mc,by,hassan1}. Also, there are some unified theories of the dark universe including dark energy and dark matter \cite{uni3,uni4,uni5,uni6, uni1, uni2}.

The idea of modifying gravity on cosmological scales has been interesting over the past
decade. In this context, recently
in the weak field limit of $f(R)$ gravity, it is shown that the corrected gravitational potential
allows estimating the total mass of a sample of 12 clusters of galaxies
 and  provides a fair fit to the mass of visible matter  estimated by X-ray observations \cite{cap}. It has been discussed that the
 Newtonian potential modified at large distances due
to the gravity propagation into the bulk \cite{flo}. The  corrections in Newtonian potential
are also parametrized by Yukawa potential \cite{long,annalen,salzano}.
The idea of modified gravity has been re-emerging with various  justifications, see for example Refs. \cite{ ec, ha,dr,rep1,rep2}. The MOND is also studied for the Milky Way and has found that inner Milky Way is completely dominated by baryon matter, which is in agreement with the predictions of standard cold dark matter (CDM) cosmology \cite{bb}. The good fits of MOND  are  found   in 15 rotation curves of low
surface brightness   galaxies \cite{wj}. In the context of MOND, the growth of inhomogeneities in a low-density is presented in the Ref. \cite{san1}. On the same track, there are  fits of samples of galaxies (mainly low surface brightness) whose dynamics are addressed  without dark matter while adopting the $f(R)$ approach \cite{cardone,salucci}. There are semi-analytical models for the disk galaxies formation in the universe dominated by dark matter, and also for the cases where the force law is given by modified Newtonian dynamics (MOND) \cite{fca}. It may be interesting to note that these models are tuned to fit the observed near-infrared Tully-Fisher relation \cite{fca}.

On the other hand, phase transitions are the fundamental property
of   interacting systems which has an important role.
From the statistical mechanics point of view, there is a connection
between the macroscopic phases and the microscopic properties of the given system.
  If there are different states in the separate regions
of multi-dimensional phase space, the transition is called first-order.
However, other transitions,  where states are
close to each other in the phase space,  are called second-order.
The effects of higher-order corrections
on the   phase transition of the clustering of particles interacting through Newton's
law have been studied in Ref.  \cite{sas2}.
Recently, the various characterization of clustering of galaxies under
the modified gravity are studied \cite{st, san, mils, sud,mir,MNRAS-Pour,PRD-Sud, MNRAS-Sud, Dark-Sud}. For instance,  a test using gas rich galaxies for which both axes of the  baryonic Tully-Fisher relation can be measured independently of the theories have been done   without the systematic uncertainty in stellar mass \cite{st}.
The effect of dark energy
  on the galactic clustering under the modified Newton's law by  cosmological
constant is studied  recently \cite{sud}.  A brane world modified Newton's gravity  in the context of  galaxy clustering   is also analysed  \cite{mir}. The motivation of this paper is
 to extend the work of Saslaw and Ahmad \cite{sas2}    by means of modified Newton's dynamics.

The zeroth-order thermodynamics of galactic clustering in the context of dark energy modification \cite{MNRAS-Pour}, and a particular modification of gravity \cite{PRD-Sud, MNRAS-Sud, Dark-Sud} have been studied already. As the second
and higher-order corrections to  thermodynamics of galaxy clustering could become important near phase transitions as clustering increases, so we want to analyse this. Now, we would like to  study the higher-order correction to the thermodynamics of system of galaxies under   modified gravity.
In this work, we study the higher-order correction to the partition function for the
gravitating system under the modified Newton's law.
In order to study the thermodynamics of galactic system, first of all we   derive the higher-order
corrected partition function. With the help resulting partition function,
we demonstrate the Helmholtz free energy, entropy, pressure, internal energy and the chemical
potential of the system. We find that
these thermodynamical quantities include corrections due to higher-order term.
A comparison with the standard expressions of the equations of state leads to the higher-order corrected clustering parameter, which characterizes the higher-order effects on
the clustering of galaxies under modified gravity.
We discuss
both the cases of the point mass and the extended mass structures of the galaxies.
The study of  extended mass structured galaxies through string theory or theory of infinitely extended particles \cite{Hes1, Hes2, Hes3} is interesting.
Under assumption that system follows quasi-equilibrium state,  we obtain a higher-order corrected distribution function for the system of galaxies.
If  the higher-order correction parameter  is turned-off, we get the
lowest-order   distribution function of the system interacting through modified gravity.
In order to study the phase transition  of clusters of galaxies, it is important
to investigate the order parameter. Here we found that the clustering parameter as an order parameter   under the modified gravity behaves very similar to  the pure Newtonian gravity case.  We find that under modified Newton's  law, the specific heat for completely
virialized system becomes negative as bound clusters dominate. Further,  the
order parameter, pressure and internal energy are formulated in terms of critical temperature.
These results  are very general which are valid for any kind of the modified gravity
as our  computations are based on the general modified gravity.

The presentation of the paper is as follows.
In section II, we discuss the higher-order expansion of two-particle function
in modified Newtonian gravity and obtain the higher-order corrected partition function.
In section III, we evaluate the more exact equation of states
corresponding to these higher-order corrections. The expression for
clustering parameter is obtained, which describes the distribution of
galaxies cluster. In the subsequent section, we invoke the corrected
distribution function which exhibits the correction and agrees with the
data. In section V, we discuss the role of clustering parameter  as
order parameter under the modified gravity circumstances. The critical temperature,  which
 maximizes specific heat,
 is evaluated in section VI.
 Final discussion with the future remarks is reported in the last section.

 \section{Higher-order corrections to  a modified gravity}
 In this section, we study the higher-order corrected partition function by considering the  gravitational system interacting under the modified Newton's gravity.
 \subsection{Thermodynamic equations}
 Our analysis is based on the important assumption that the clustering of the galaxies (as gravitating particles) in the expanding universe evolve in the quasi-equilibrium manner, which is a consequence
 of equilibrium states, and thus forms an ensemble of comoving cells. For such a large system,
 the all cells of the ensemble are of the equal
volume $V$ and equal average density.
According to the standard classical statistical
mechanics, the partition function of a system of $N$  particles (galaxies)  of mass $m$, interacting with a  gravitational  potential energy
$\Phi$,  and average temperature $T$ is given by \cite{ahm02}:
\begin{equation}\label{zn}
Z_N(T,V)=\frac{1}{N!}\left(\frac{2\pi mT}{\lambda^2}\right)^{3N/2}Q_N(T,V),
\end{equation}
where the configurational integral, $Q_{N}(T,V)$, in terms of two-particle function $f_{ij}$ can be expressed as:
\begin{equation}\label{q2}
Q_{N}(T,V)=\int....\int \biggl[(1+f_{12})(1+f_{13})(1+f_{23})(1+f_{14})\dots (1+f_{N-1,N})\biggr]d^{3}r_{1}d^{3}r_{2}\dots d^{3}r_{N}.
\end{equation}
The two-particle function and gravitational  potential energy have following relation:
\begin{equation}\label{fun}
f_{ij}=e^{-\Phi(r_{ij})/T}-1.
\end{equation}
We note that the two-particle function exists only if  interactions are present there in the system.

We are interested in modified Newtonian gravity here because MOND paradigm  can boast of some
successful predictions regarding galactic dynamics.
The departure of  gravitation   from Newtonian theory due to MOND should be inside the
virial radius, where dynamical accelerations are negligibly small from the perspective to MOND.
The   gravitational potential modified by a  sufficiently regular function ${\mathcal{F}}(r_{ij})$  \cite{san} is given by
\begin{equation}\label{phi}
\Phi(r_{ij})=-\frac{Gm^2}{\sqrt{r_{ij}^{2}+\epsilon^{2}}}- {\mathcal{F}}(r_{ij},\epsilon),
\end{equation}
where a softening parameter $\epsilon$ is used to take care of divergences caused in Hamiltonian. Also, force ${\mathcal{F}}(r_{ij},\epsilon)$ is infinitesimal at large distance.
By exploiting Eqs. (\ref{fun}) and  (\ref{phi}),  the two-particle function
can be expressed as following:
\begin{equation}
f_{ij}=\exp\left[\frac{Gm^{2}}{T\sqrt{r_{ij}^{2}+\epsilon^{2}} }+
\frac{{\mathcal{F}}(r_{ij},\epsilon)}{T}\right]-1.
\end{equation}
By expanding exponential term, this further simplifies to
\begin{eqnarray}
f_{ij}=\left(\frac{Gm^{2}}{T\sqrt{r_{ij}^{2}+\epsilon^{2}}}+
\frac{{\mathcal{F}}(r_{ij},\epsilon)}{T}\right) + \frac{1}{2!} \left(
\frac{Gm^{2}}{T\sqrt{r_{ij}^{2}+\epsilon^{2}}}+
\frac{{\mathcal{F}}(r_{ij},\epsilon)}{T}\right)^2 + ... .
\end{eqnarray}
Following the standard technique, we evaluate the
 the configuration integral  (\ref{q2}) for above two-particle function  over a spherical volume of radius $R_{1}$  for $N = 1$  as follows,
\begin{equation}
Q_{1}(T,V)=V.
\end{equation}
For $N = 2$, the configuration integral $Q_{2}(T,V)$  has the following expression:
\begin{equation}
Q_{2}(T,V)=4\pi V \int_{0}^{R}\left[1+\frac{Gm^2}{T\sqrt{r^{2}+\epsilon^{2}}}+\frac{{\mathcal{F}}(r,\epsilon)}{T}   + \frac{1}{2!} \left(
\frac{Gm^{2}}{T\sqrt{r^{2}+\epsilon^{2}}}+
\frac{{\mathcal{F}}(r,\epsilon)}{T}\right)^2 \right]r^2dr.
\end{equation}
This further simplifies to,
\begin{eqnarray}\label{a}
Q_{2}(T,V)&=&V^2\left[1+  \frac{3}{2}\frac{Gm^2}{R T}\left( \sqrt{1+\frac{\epsilon^2}{R ^2}} + \frac{\epsilon^2}{R ^2} \log \frac{\epsilon/R }{\left[ 1+\sqrt{1+\frac{\epsilon^2}{R ^2}}\right]}+\frac{2}{Gm^2R^2}\int_0^R  {\mathcal{F}}(r,\epsilon)  r^2 dr\right)  \right.\nonumber\\
&+&\left.  \frac{3}{2}\left(\frac{Gm^2}{R T}\right)^2 \left[1-\frac{\epsilon}{R}
\tan^{-1} \left(\frac{R}{\epsilon} \right)+\frac{2}{3}\frac{R^2}{Gm^2}\int_0^R
  \frac{{\mathcal{F}}(r,\epsilon)}{\sqrt{r^{2}+\epsilon^{2}}}
  r^2 dr \right. \right.\nonumber\\
&+&\left.\left.\frac{1}{3 }\left(\frac{R}{Gm^2}\right)^2\int_0^R  {\mathcal{F}}^2(r,\epsilon)r^2 dr \right] \right].
\end{eqnarray}
We rewrite the equation (\ref{a}) as follows,
\begin{equation}
Q_{2}(T,V)=V^2\big(1+\alpha_1 x +\alpha_2 x^2  \big),
\end{equation}
where the   following definitions for dimensionless parameters have been utilized:
\begin{eqnarray}
&&\alpha_1  = \sqrt{1+\frac{\epsilon^2}{R ^2}} + \frac{\epsilon^2}{R ^2} \log \frac{\epsilon/R }{\left[ 1+\sqrt{1+\frac{\epsilon^2}{R ^2}}\right]}+\frac{2}{Gm^2R^2}\int_0^R  {\mathcal{F}}(r,\epsilon)  r^2 dr,\label{para}\\
&&
  \alpha_2= \frac{2}{3}\left[1-\frac{\epsilon}{R}
\tan^{-1} \left(\frac{R}{\epsilon} \right)+\frac{2}{3}\frac{R^2}{Gm^2}\int_0^R
  \frac{{\mathcal{F}}(r,\epsilon)}{\sqrt{r^{2}+\epsilon^{2}}}
  r^2 dr +\frac{1}{3 }\left(\frac{R}{Gm^2}\right)^2\int_0^R  {\mathcal{F}}^2(r,\epsilon)r^2 dr\right],\label{para1}\\
&&x= \frac{3}{2}\left(\frac{ Gm^{2}}{ R T}\right)^3=\frac{3}{2}\left(\frac{ Gm^{2}}{  T}\right)^3\rho.\label{sc}
\end{eqnarray}
The last definition utilizes the
  scale transformations  $\rho\rightarrow \lambda^{-3}\rho, T\rightarrow\lambda^{-1}T$ and $R \rightarrow \lambda R $,
to transform $\frac{Gm^2}{R T}\rightarrow (\frac{Gm^2}{R T})^3$ and $R_{1}\sim \rho^{-1/3}\sim (\bar N/V)^{-1/3}$.

In case of point mass galaxies, the  dimensionless parameters (\ref{para}) and (\ref{para1}) reduce to
\begin{eqnarray}\label{alp}
&&\alpha_1|_{\epsilon=0}  = 1 +  \frac{2}{Gm^2R^2}\int_0^R  {\mathcal{F}}(r,0)  r^2 dr, \\
&&
  \alpha_2|_{\epsilon=0}= \frac{2}{3}\left[1  +\frac{2}{3}\frac{R^2}{Gm^2}\int_0^R
  {{\mathcal{F}}(r,0)}
  r  dr +\frac{1}{3 }\left(\frac{R}{Gm^2}\right)^2\int_0^R  {\mathcal{F}}^2(r,0)r^2 dr\right].
\end{eqnarray}
The    configuration integral for $N$ particles case can be obtained  recursively
as following:
\begin{equation}\label{qn}
Q_{N}(T,V)=V^N\big(1+\alpha_1 x +\alpha_2 x^2  \big)^{N-1}.
\end{equation}
Plugging this resulting $N$ particles configuration integral in to  (\ref{zn}), we obtain the
explicit expression of gravitational partition function as
\begin{equation}\label{part}
Z_N(T,V)=\frac{1}{N!}\left(\frac{2\pi mT}{\lambda^2}\right)^{3N/2}V^{N}\big(1+\alpha_1 x +\alpha_2 x^2  \big)^{N-1}.
\end{equation}
This is the higher-order corrected partition function of $N$ particles (galaxies)
interacting under modified Newtonian potential.
\section{Equations of State}
The gravitational system (galaxies cluster) can be described adequately by macroscopic thermodynamic
variables, such as total energy ($U$), Helmholtz free energy ($F$), Gibbs free energy ($G$), enthalpy ($H$)
 entropy ($S$), temperature ($T$),
pressure ($P$), volume ($V$), number of particles (galaxies) ($N$) and chemical
potential ($\mu$).
We compute  these thermodynamic quantities for the infinite statistically homogeneous
system of gravitating particles which are characterized by exact equations of state.
As we have   expression for the gravitational partition function (\ref{part}),
  these thermodynamic quantities can  be easily calculated.
For instance, utilizing relation $F=-T\ln Z_{N}(T,V)$, the Helmholtz free energy
 is demonstrated  as
\begin{equation}\label{f}
F=-T\ln\biggl[\frac{1}{N!}\left(\frac{2\pi mT}{\lambda^2}\right)^{3N/2}V^N\big(1+\alpha_1 x +\alpha_2 x^2  \big)^{N-1}\biggr].
\end{equation}
The Helmholtz free energy is nothing but the thermodynamic potential that measures the   work obtainable  from a   thermodynamic system at a constant temperature. The negative of the difference in the
Helmholtz energy gives information regarding  the maximum amount of work that the system
can perform in a thermodynamic process in which volume is held constant.
Further simplification leads to,
\begin{equation}\label{hem}
F=NT\ln\left(\frac{N}{V}T^{-3/2}\right)-NT -(N-1)T\ln\big(1+\alpha_1 x +\alpha_2 x^2  \big) -\frac{3}{2}NT\ln\left(\frac{2\pi m}{\lambda^2}\right).
\end{equation}

\begin{figure}[htb]
 \begin{center}
 \begin{tabular}{cc}
 {\resizebox{90mm}{!}{ \includegraphics[width=200pt]{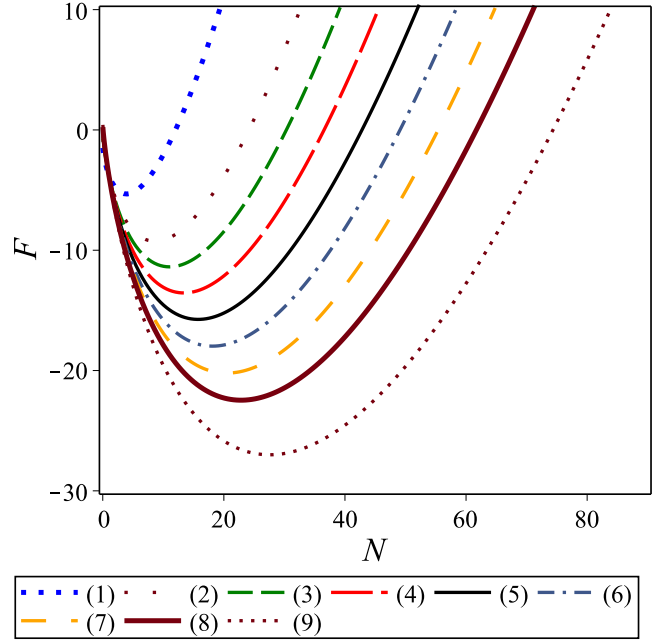}}}
\end{tabular}
 \end{center}
 \caption{Typical behavior of Helmholtz free energy in terms of particle number with variation of $\alpha_{1}$ and $\alpha_{2}$ and unit value for other parameters. (1) $\alpha_{1}<0$, $\alpha_{2}<0$;\hspace{2mm} (2) $\alpha_{1}=0$, $\alpha_{2}<0$;\hspace{2mm} (3) $\alpha_{1}<0$, $\alpha_{2}=0$;\hspace{2mm} (4) $\alpha_{1}>0$, $\alpha_{2}<0$;\hspace{2mm} (5) $\alpha_{1}=0$, $\alpha_{2}=0$;\hspace{2mm} (6) $\alpha_{1}<0$, $\alpha_{2}>0$;\hspace{2mm} (7) $\alpha_{1}>0$, $\alpha_{2}=0$;\hspace{2mm} (8) $\alpha_{1}=0$, $\alpha_{2}>0$;\hspace{2mm} (9) $\alpha_{1}>0$, $\alpha_{2}>0$.}
  \label{fig1}
 \end{figure}

Although $\alpha_{1}$ and $\alpha_{2}$  depend on ${{\mathcal{F}}(r,\epsilon)}$, but  can be fix as they are independent of $N$, $V$ and $T$. In that case, we can obtain graphical behavior of the Helmholtz free energy by varying $\alpha_{1}$ and $\alpha_{2}$.
In the Fig. \ref{fig1}, we can see the behavior of Helmholtz free energy with respect to $N$. We can see a minimum which is corresponding to the maximum of distribution function calculated later. There exists  critical $N$ also  where Helmholtz free energy vanishes, which means that the cluster of galaxies is at equilibrium. For  large $N$,  we can see a divergence in the Helmholtz free energy which yields to the state out of equilibrium. The last line, which is drawn for $\alpha_{1}>0$ and $\alpha_{2}>0$, may correspond to the case of ${{\mathcal{F}}(r,\epsilon)}=0$. It means that the effect of higher-order correction is a decreasing potential.

The next step is to derive entropy in terms of the model parameters. The entropy and the free energy are related by  following expression:
$S= -\biggl(\frac{\partial F}{\partial T}\biggr)_{N,V}$.
Therefore, for a given Helmholtz free energy (\ref{hem}),
the entropy reads
\begin{equation}
S=N\ln\left(\frac{V}{N}T^{3/2}\right)+(N-1)\ln\big(1+\alpha_1 x +\alpha_2 x^2  \big)-3
(N-1)\frac{\alpha_1 x +2\alpha_2 x^2}{ 1+\alpha_1 x +\alpha_2 x^2   }+\frac{5}{2}N+\frac{3}{2}N\ln\left(\frac{2\pi m}{\lambda^2}\right).\label{s}
\end{equation}
 For the large $N$, we can use  $N-1\approx N$   approximation  and, thus,
the specific entropy  is simplified to
\begin{equation}
\frac{S}{N}=\ln\left(\frac{V}{N}T^{3/2}\right)-\ln\left(1-\frac{ \alpha_1 x +\alpha_2 x^2}{1+\alpha_1 x +\alpha_2 x^2}\right)-3\frac{ \alpha_1 x +2\alpha_2 x^2}{1+\alpha_1 x +\alpha_2 x^2}+\frac{S_{0}}{N},
\end{equation}
where    $S_{0}=\frac{5}{2}N+\frac{3}{2}N\ln\left(\frac{2\pi m}{\lambda^2}\right)$ is
the fiducial entropy. Here we notice that, for $\alpha_1=\alpha_2=0$, this reduces to
the perfect classical gas case.

The internal energy of a gravitational system is the energy contained within the system, including the kinetic and potential energy as a whole. For given state of a system it cannot be  measured directly. However, once you know the  state variables of the system,
free energy $F$, temperature $T$ and entropy  $S$,  the internal energy can easily be calculated through relation, $U =  F+TS$. This follows,
\begin{equation}
U=\frac{3}{2}NT\left(\frac{1- \alpha_1 x -3\alpha_2 x^2}{1+\alpha_1 x +\alpha_2 x^2}\right).
\end{equation}
We observe here to, that the perturbation from perfect classical gas is due to
non-vanishing parameters till higher-order  $\alpha_1$ and $\alpha_2$.

\begin{figure}[htb]
 \begin{center}
 \begin{tabular}{cc}
 {\resizebox{90mm}{!}{ \includegraphics[width=200pt]{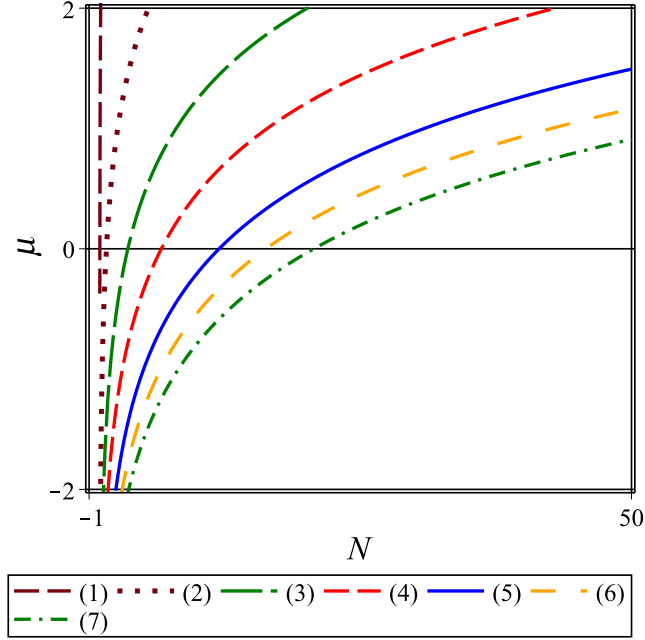}}}
\end{tabular}
 \end{center}
 \caption{Typical behavior of Chemical potential in terms of particle number with variation of $\alpha_{1}$ and $\alpha_{2}$, we set unit value for other parameters. (1) $\alpha_{1}<0$, $\alpha_{2}<0$;\hspace{2mm} (2) $\alpha_{1}=0$, $\alpha_{2}<0$;\hspace{2mm} (3) $\alpha_{1}<0$, $\alpha_{2}=0$;\hspace{2mm} (4) $\alpha_{1}=0$, $\alpha_{2}=0$;\hspace{2mm} (5) $\alpha_{1}<0$, $\alpha_{2}>0$;\hspace{2mm} (6) $\alpha_{1}=0$, $\alpha_{2}>0$;\hspace{2mm} (7) $\alpha_{1}>0$, $\alpha_{2}>0$.}
  \label{fig2}
 \end{figure}

By the relation $P= -\left(\frac{\partial F}{\partial V}\right)_{N,T}$, it is straightforward to calculate pressure equation of state, which is given by
\begin{eqnarray}\label{p}
  P=\frac{NT}{V}\left(\frac{1-\alpha_2 x^2}{{1+\alpha_1 x +\alpha_2 x^2}}\right).
\end{eqnarray}
Then, by using the relation $G=F+PV$, one can obtain Gibbs free energy which behaves as the Helmholtz free energy.
 Chemical potential $\mu$ gives information about the change in internal energy if
 one particle is added to the system, while keeping all other thermodynamic quantities
 constant. In order to derive   chemical potential, we exploit the relation  $\mu = \biggl(\frac{\partial F}{\partial N}\biggr)_{V,T}$  and get
  \begin{eqnarray}\label{mu}
 {\mu}={T}\ln\left(\frac{N}{V} T^{-3/2}\right)-{T}\ln\big( 1+\alpha_1 x +\alpha_2 x^2\big)-\frac{3}{2}{T}\ln\left(\frac{2\pi m}{\lambda^2}\right)-{T}\frac{ \alpha_1 x +2\alpha_2 x^2}{1+\alpha_1 x +\alpha_2 x^2}.
 \end{eqnarray}
 In  the above expression of chemical potential,  an extra term of order $x^2$ is present which
(physically) represents the energy required to put in a galaxy to
under-virialized clustering regions and this, of course, will affect the distribution
functions.
In the Fig. \ref{fig2}, we can see a typical behavior of the chemical potential with $N$ for different values of $\alpha_{1}$ and $\alpha_{2}$. There exist  some critical $N$ also where $\mu=0$, which means that $\frac{\partial S}{\partial N}=0$. It is clear that as $N$ increases (decreases) $\mu$ increases (decreases). More precisely, if the number
density $\frac{N}{V}$ increases (decreases) with fixed $T$, the chemical potential increases (decreases). The last line, which is dash dot green line, may corresponds to the case of
 ${{\mathcal{F}}(r,\epsilon)}=0$. It tells us that the higher-order correction is increasing  the chemical potentials.

Equating above expressions of thermodynamic quantities to their  standard forms
given in terms of clustering parameter (for details see, e.g., \cite{ahm02}),
the value of higher-order corrected clustering parameter,    $\mathcal{B}$,
   for the system of galaxies  in the expanding universe emerges, naturally, as
\begin{eqnarray}
\mathcal{B}=\frac{ \alpha_1 x +2\alpha_2 x^2}{1+\alpha_1 x +\alpha_2 x^2}.
\label{b}
\end{eqnarray}
It should be noted that the clustering parameter plays an important role in the clustering of galaxies.
For the point mass case  (  i.e., $\epsilon =0$),  the higher-order corrected clustering parameter  in modified potential energy takes following value:
\begin{eqnarray}
\mathcal{B}_0  =\frac{\left[ 1 +  \frac{2}{Gm^2R^2}\int_0^R  {\mathcal{F}}(r,0)  r^2 dr\right] x+\frac{4}{3}\left[1  +\frac{2}{3}\frac{R^2}{Gm^2}\int_0^R
  {{\mathcal{F}}(r,0)}
  r  dr +\frac{1}{3 }\left(\frac{R}{Gm^2}\right)^2\int_0^R  {\mathcal{F}}^2(r,0)r^2 dr\right]x^2}{1+\left[ 1 +  \frac{2}{Gm^2R^2}\int_0^R  {\mathcal{F}}(r,0)  r^2 dr\right] x+\frac{2}{3}\left[1  +\frac{2}{3}\frac{R^2}{Gm^2}\int_0^R
  {{\mathcal{F}}(r,0)}
  r  dr +\frac{1}{3 }\left(\frac{R}{Gm^2}\right)^2\int_0^R  {\mathcal{F}}^2(r,0)r^2 dr\right]x^2}.
\end{eqnarray}
Here we notice that for Newtonian   potential energy, i.e., $\mathcal{F}(r)=0$,
 the higher-order corrected clustering parameter has the following form:
 \begin{eqnarray}
\mathcal{B}_0  =\frac{ x+\frac{4}{3} x^2}{1+  x+\frac{2}{3} x^2}.
\end{eqnarray}
This coincides with the original clustering parameter $b=\frac{x}{1+x}$ for
galaxies with point mass structure when the higher-order corrections are switched off.
\section{Corrected Distribution Function}
In order to obtain the higher-order  corrected gravitational quasi-equilibrium
distribution function $f(N)$, we  follow the method discussed in Ref. \cite{sas1}.
We consider a grand canonical ensemble of cells with the same shape and volume, which are much smaller than the total gravitational system. Here, the number of galaxies and their
mutual gravitational energy varies among cells.
The probability of finding a particular number of galaxies in a cell of volume $V$
follows from summation over all energy states, i.e.,
\begin{eqnarray}\label{fn}
p(N)=\sum_i \Omega_{Ni},
\end{eqnarray}
where $\Omega_{N_i}$ is the probability of finding $N_i$ particles in the energy states $U_i(N_i,V)$ with the following expressions:
\begin{eqnarray}
\Omega_{Ni}=\frac{e^{ N_i\mu /T -U_i/T}}{{Z_G(T,V,\mu)}}
\end{eqnarray}

Here, the grand canonical partition function has following expression:
\begin{eqnarray}
Z_{G}(T,V,\mu)=\sum_{i} e^{\frac{N_i\mu -U_i}{T}}=e^{-\frac{\Psi}{T}}, \label{g}
\end{eqnarray}
where $\Psi$, so-called the grand canonical potential, is nothing but the Legendre transformation of the average energy concerning both $S$ and $N$. Hence, this can be given by $\Psi =-PV$.
Now, by exploiting relation  (\ref{p}), the higher-order corrected  grand partition function (\ref{g}) for the galaxies interacting through modified gravity
  is
calculated by
\begin{equation}
 Z_{G}=e^{\frac{PV}{T}}=  e^{\bar N(1-\mathcal{B})}.\label{z}
\end{equation}

\begin{figure}[htb]
 \begin{center}
 \begin{tabular}{cc}
 {\resizebox{80mm}{!}{ \includegraphics[width=200pt]{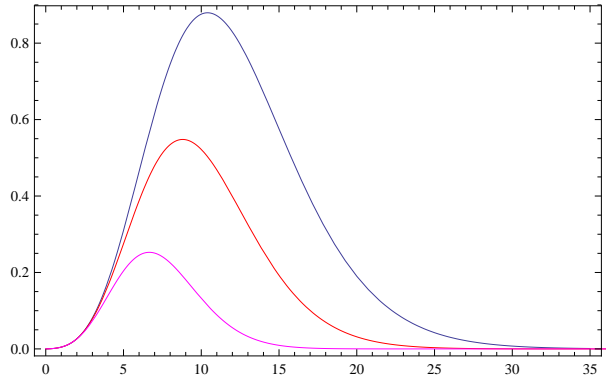}}}
\end{tabular}
 \end{center}
 \caption{Comparative study of distribution function $p(N,\epsilon)$ (perpendicular axis) versus $N$
 (horizontal axis) for extended mass structure for $\mathcal{B} =0.3$,
 and $\bar{N}=10$. Here,  $\alpha_2  =0$ (Violet bottom line) corresponds to no higher-order corrections, for $\alpha_2   =2/5, \alpha_1 =\sqrt{2}$  (Red middle line)  and  $\alpha_2   =4/3, \alpha_1 =\sqrt{2}$  (Magenta top line) denote   higher-order   corrections.}
 \label{fig3}
 \end{figure}

The probability of finding $N$ particles in a cell of volume $V$ is given by
\begin{eqnarray}
p(N,\epsilon)= e^{\frac{N\mu}{T}}\left(\sum_{i}e^{\frac{-U_i}{T}}\right)e^{-\bar N(1-\mathcal{B})}= {e^{\frac{N\mu}{T}}Z_{N}(V,T)}e^{-\bar N(1-\mathcal{B})}.\label{dis}
\end{eqnarray}
 Upon solving the quadratic equation (\ref{b}) for $x$, we get
 \begin{eqnarray}
 \alpha_1 x+\alpha_2 x^2 =\frac{\mathcal{B}}{1-\mathcal{B}}-\frac{\mathcal{B}^2(2-\mathcal{B})}{(1-\mathcal{B})^3}\frac{\alpha_2}{\alpha_1^2}.
 \end{eqnarray}
 Now, with the help of Eqs. (\ref{part}) and (\ref{mu}) together with (\ref{alp}), the distribution function (\ref{dis}) for extended mass
particles (galaxies with halos) is demonstrated as
\begin{equation}
p(N,\epsilon)=\frac{\bar{N}^{N}}{N!}\biggl(1+\frac{N}{\bar N}\frac{\mathcal{B}}{(1-\mathcal{B})} - \frac{N}{\bar N}\frac{\mathcal{B}^2(2-\mathcal{B})}{(1-\mathcal{B})^3}
\frac{\alpha_2}{\alpha_1^2}\biggr)^{N-1}\biggl(1+\frac{\mathcal{B}}{(1-\mathcal{B})}-\frac{\mathcal{B}^2(2-\mathcal{B})}{(1-\mathcal{B})^3}
\frac{\alpha_2}{\alpha_1^2}\biggr)^{-N}e^{-N\mathcal{B} -\bar N(1-\mathcal{B})}.
\end{equation}
By setting $\epsilon=0$,  we can get the gravitational quasi-equilibrium distribution function
for  galaxies with point mass structures as follows,
\begin{equation}
p(N,0)=\frac{\bar{N}^{N}}{N!}\biggl(1+\frac{N}{\bar N}\frac{\mathcal{B}_0}{(1-\mathcal{B}_0)} - \frac{N}{\bar N}\frac{\mathcal{B}_0^2(2-\mathcal{B}_0)}{(1-\mathcal{B}_0)^3}
\frac{\alpha_2}{\alpha_1^2}\biggr)^{N-1}\biggl(1+\frac{\mathcal{B}_0}{(1-\mathcal{B}_0)}-\frac{\mathcal{B}_0^2(2-\mathcal{B}_0)}{(1-\mathcal{B}_0)^3}
\frac{\alpha_2}{\alpha_1^2}\biggr)^{-N}e^{-N\mathcal{B}_0 -\bar N(1-\mathcal{B}_0)}.
\end{equation}
This result  matches exactly with the result obtained in Ref. \cite{sas2} except with modified
clustering parameters.
 The presence of higher-order correction is obvious here as this expression contains
 $\alpha_2$.
 For $\alpha_2\rightarrow 0$, we can get the lowest order gravitational quasi-equilibrium
 distribution function with modified gravity. Further, for $\mathcal{F}(r)=0$, this reduces to
 original result discussed in \cite{ahm02}.\\
From the Fig. \ref{fig3}, we can compare the behavior of distribution function with and without higher-order corrections. Here, it is clear that the presence of higher-order correction
 increases the value of the distribution function. Also, there exists a critical $N$ which maximizes the probability and corresponds to the minimum of free energies.

\section{Order Parameter for Phase Transition}
In order to study the phase transition for galaxies evolution and cluster formation, it is important to analyze the order parameter which characterizes the phase transition.
In this regard, let us write the
clustering parameter for point mass galaxies under modified potential to lowest order
as an order parameter
\begin{eqnarray}
 B_0  =\frac{\left( 1 +  \frac{2}{Gm^2R^2}\int_0^R  {\mathcal{F}}(r,0)  r^2 dr\right) x }{1+\left( 1 +  \frac{2}{Gm^2R^2}\int_0^R  {\mathcal{F}}(r,0)  r^2 dr\right) x }.
\end{eqnarray}
An extensive quantity, so-called conjugate field, corresponding to the parameter $B_0$ is defined by
\begin{eqnarray}
\mathcal{H}(T,V,N)  =\frac{NT}{B_0}=-NT\left[\frac{1+\left( 1 +  \frac{2}{Gm^2R^2}\int_0^R  {\mathcal{F}}(r,0)  r^2 dr\right) x }{\left( 1 +  \frac{2}{Gm^2R^2}\int_0^R  {\mathcal{F}}(r,0)  r^2 dr\right) x }\right].
\end{eqnarray}
This is justified by considering the work done during adiabatic expansion.
From the above expression, one can see that even in modified gravity this behaves similar to the unmodified gravity \cite{sas2}, i.e., as $T\rightarrow 0$, $B_0\rightarrow 1$ and $\mathcal{H}\rightarrow 0$; however as $T\rightarrow \infty$, $B_0\rightarrow 0$ and $\mathcal{H}\rightarrow -\infty$.

Now,  the  change in  $\mathcal{H}$ with temperature $T$ at constant volume $V$ and constant
$N$ reads,
\begin{eqnarray}
\left(\frac{\partial \mathcal{H}}{\partial T}\right)_{V,N} &=&-\frac{N}{B_0}-\frac{2T^3 V}{(Gm^2)^3\left( 1 +  \frac{2}{Gm^2R^2}\int_0^R  {\mathcal{F}}(r,0)  r^2 dr\right)},\nonumber\\
&=& -\frac{8VT^3}{3(Gm^2)^3\left( 1 +  \frac{2}{Gm^2R^2}\int_0^R  {\mathcal{F}}(r,0)  r^2 dr\right)} -N.
\end{eqnarray}
With respect to volume $V$, it varies as
\begin{eqnarray}
\left(\frac{\partial \mathcal{H}}{\partial V}\right)_{T,N} &=&- \frac{2T^4}{3G^3m^6 \left(1 +  \frac{2}{Gm^2R^2}\int_0^R  {\mathcal{F}}(r,0)  r^2 dr \right)},
\end{eqnarray}
and with respect to $N$, it varies as
\begin{eqnarray}
\left(\frac{\partial \mathcal{H}}{\partial N}\right)_{T,V} &=&-T.
\end{eqnarray}
From the above expressions, it is obvious that although these derivatives
are different to  those derived in
\cite{sas2} but have similar behavior, i.e., they diverge
at $T\rightarrow \infty$. However,   $T\rightarrow 0$ corresponds to $B_0\rightarrow 0$,
so $B_0=0$ can be considered as a critical point. At this point, we remark that the order parameter
and conjugate field of galaxies cluster
in modified gravity also behave in similar fashion to that  Newtonian gravity case.

\section{Critical Temperature}
\begin{figure}[htb]
 \begin{center}
 \begin{tabular}{cc}
 {\resizebox{90mm}{!}{ \includegraphics[width=200pt]{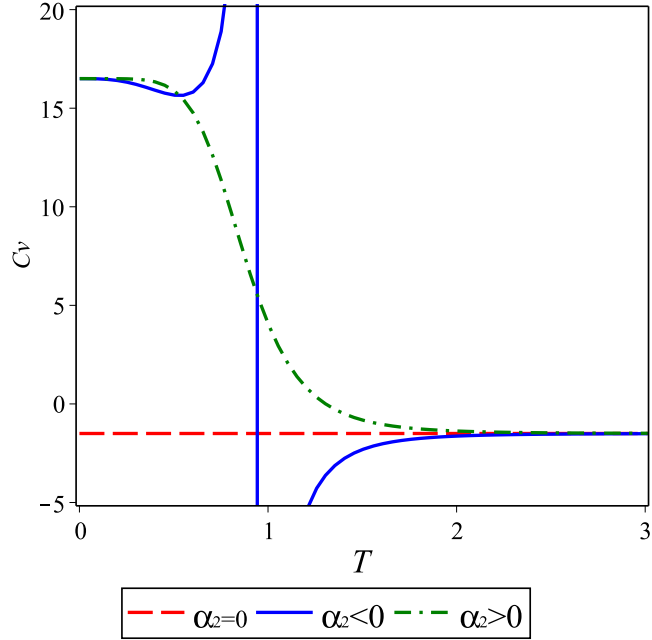}}}
\end{tabular}
 \end{center}
 \caption{Typical behavior of specific heat in terms of temperature for the case of $\mathcal{B} = 1$ with variation of $\alpha_{2}$. We set unit value for other parameters. $\alpha_{2}<0$ (blue solid); $\alpha_{2}=0$ (red dashed); $\alpha_{2}>0$ (dash dot green).}
  \label{fig4}
 \end{figure}

The variations of the specific heat, from  perfect gas  ($\mathcal{B} = 0$) to  fully
virialized gas  ($\mathcal{B} = 1$), provide illuminating physical insights into clustering.
The higher-order corrected specific heat at constant volume is calculated as
\begin{eqnarray}
C_V =\frac{1}{N}\left(\frac{\partial U}{\partial T}\right)_{N,V}=\frac{3}{2}\left[1+4\mathcal{B} -6\mathcal{B}^2 +\frac{12\alpha_2 x^2}{(1+\alpha_1 x+\alpha_2 x^2)} \right].
\end{eqnarray}
For $\alpha_1,\alpha_2\rightarrow 0$ ($\mathcal{B}\rightarrow0$), the specific heat at constant volume ($C_V$) is $3/2$, which corresponds to a monotonic perfect gas.
For completely virialized system (i.e., $\mathcal{B} = 1$), the value of specific heat is
\begin{eqnarray}
C_V =  \frac{18\alpha_2 x^2}{(1+\alpha_1 x+\alpha_2 x^2)} - \frac{3}{2}.
\end{eqnarray}
It is well-known that  the negative specific heat corresponds to the instability  in the system. In general, the nature of these instabilities of gravitationally interacting systems is not  same   to imperfect gases because gravitationally interacting systems add  extra degrees of freedom due to semi-stability. In this situation, many galaxies   join  clusters and result additional energy which, on average,  rises out the cluster potential wells. As a result, they lose kinetic energy and cool, thus producing the overall negative  value of specific heat. In the Fig. \ref{fig4}, we can see the behavior of specific heat with the variation of $\alpha_{1}$ and $\alpha_{2}$. From the figure, we observe that the variation of $\alpha_{1}$ is not much important while variation of $\alpha_{2}$ has an important effect. It can yield to positive specific heat for example by setting $\alpha_{2}=1$. Also $\alpha_{2}<0$ may lead  to the phase transition which is illustrated by divergence of solid blue  line in the Fig. \ref{fig4}.

 However, for  $\mathcal{B} = 1/3$, the specific heat becomes
  \begin{eqnarray}
C_V =   \frac{5}{2}+ \frac{18\alpha_2 x^2}{(1+\alpha_1 x+\alpha_2 x^2)},
\end{eqnarray}
which describes a higher-order corrected behavior of diatomic gas.
 The critical temperature can be obtained by
maximizing $C_V$ as following:
\begin{eqnarray}
\frac{\partial C_V}{\partial T} =0.
\end{eqnarray}
For $\alpha_2 =0$, this  yields to
\begin{eqnarray}
2\alpha_1 x =1.
\end{eqnarray}
For point masses galaxies, the critical temperature is given by
\begin{eqnarray}
T_c =\left[3(Gm^2)^3\frac{\bar N}{V} +6\left(\frac{Gm^2}{R}\right)^2\frac{\bar N}{V}{\int_0^R  {\mathcal{F}}(r,0)  r^2 dr}\right]^{1/3}.
\end{eqnarray}
In case of extended mass galaxies,  the critical temperature is given by
 \begin{eqnarray}
T_c =\left[3(Gm^2)^3\frac{\bar N}{V} \alpha_1\right]^{1/3},
\end{eqnarray}
where $\alpha_1$ is given in (\ref{alp}).

For both point mass and extended mass cases, the specific heat in terms of critical temperature  is expressed as
\begin{eqnarray}
C_V =\frac{3}{2}\left[1-2\frac{1-4\left(\frac{T}{T_c}\right)^3}{\left(1+2\left(\frac{T}{T_c}\right)^3 \right)^2}\right],
\end{eqnarray}
which matches with the expression calculated in Ref. \cite{sas2}.
The above justifies that at critical temperature,
the basic homogeneity of the
system may break on the average of interparticle scale
which has been caused by the formation of binary gravitational systems.
The clustering parameter in terms of critical temperature is given by
\begin{eqnarray}
\mathcal{B}=\frac{T_c^3}{T_c^3+2 {T}^3 }.
\end{eqnarray}
This shows that  at $T=T_c$ the (critical) corrected clustering parameter ($\mathcal{B}_{crit}$) is $1/3$, at which   the specific heat takes maximum value.

The   pressure    and internal energy,  in terms of critical temperature, are
written by
 \begin{eqnarray}
&& P=\frac{2N}{V}\left(\frac{T^4  }{ T_c^3+2  {T}^3 }\right), \\
  &&{U}= \frac{3}{2}NT\left(\frac{2T^3-T_c^3  }{ 2T^3+T_c^3  }\right).
 \end{eqnarray}
 Clearly, at $T=T_c$, the value of pressure and internal energy become
 $P=\frac{2}{3}\frac{NT}{V}$ and ${U}= \frac{1}{2}NT$.
\section{Discussion and Conclusion}
In this paper, we have investigated the higher-order correction to
 the partition function for the gravitational system interacting through
 modified gravity.  We have computed all the results corresponding to  modified Newtonian gravity with general correction term. It should be remembered  that this model  is  neither suitable for the case where   force is a function of the velocity of the particles nor for the case of anisotropic universe. By developing
  the higher-order corrected partition function, we
studied  the thermodynamics of  galactic system. In particular,
we have derived the Helmholtz free energy,  Gibbs free energy, entropy, enthalpy, pressure,
internal energy and the chemical potential. All of these
thermodynamical quantities get correction due to
the higher-order term.
By comparing the expressions of these
thermodynamical quantities to their standard forms, we have obtained the
higher-order corrected clustering parameter which characterizes the clustering
of galaxies  under modified gravity. Within this context, we
 have discussed both the point mass and extended mass structures of galaxies.
Further, we have discussed the higher-order corrected distribution function
for this system. By switching-off the
higher-order correction parameter,  one gets the
lowest order gravitational quasi-equilibrium distribution function corresponding to modified gravity.

In order to study the phase transition for galaxies evolution and cluster formation, it is important to investigate the
order parameter.  The behaviors of order parameter  together with the conjugate
field for galaxy clusters in modified gravity  have found similar to that of pure Newtonian
gravity case. The clustering parameter as an order parameter is related to the
(so-called) critical temperature at which the specific heat takes maximum value.  We have found that for modified gravitational
interaction   the specific heat for completely virialized system also becomes
negative because bound clusters dominate. Further, we have formulated the
order parameter, internal energy  and the pressure in terms of the critical temperature.

\begin{acknowledgments}
S.C. acknowledges the support of INFN ({\it iniziative specifiche}
TEONGRAV and QGSKY) and  the COST Action CA15117
(CANTATA), supported by COST (European Cooperation in Science and
Technology).
\end{acknowledgments}

\end{document}